\documentclass[prl,twocolumn, superscriptaddress, preprintnumbers]{revtex4}

\usepackage{amsfonts,amssymb,amsmath}
\usepackage{bm,bbm}
\usepackage{graphicx, epsfig}

\newcommand{\be}{\begin{equation}}
\newcommand{\ee}{\end{equation}}
\newcommand{\bea}{\begin{eqnarray}}
\newcommand{\eea}{\end{eqnarray}}

\newcommand{\stopR}{\tilde{t}_R}
\newcommand{\st}{\tilde{t}}

\newcommand{\GeV}{~\text{GeV}}
\newcommand{\invfb}{~\text{fb}^{-1}}

\begin{document}
\preprint{RUNHETC-2011-10}

\title{Measuring Top Squark Interactions With The Standard Model Through Associated Production}

\author{Spencer Chang}
\affiliation{Physics Department, University of California Davis, Davis, California 95616}
\affiliation{Department of Physics, University of Oregon, Eugene, OR 97403}

\author{Can Kilic}
\affiliation{Department of Physics and Astronomy, Rutgers University, Piscataway, NJ 08854, USA}

\author{Takemichi Okui}
\affiliation{Department of Physics, Florida State University, Tallahassee, FL 32306}

\begin{abstract}
A new particle's interactions can be measured at colliders, by observing its associated production with Standard Model particles.  We focus on the case of a collider-stable right-handed top squark and study the LHC sensitivities to its couplings to the photon, $Z$, and the Higgs boson.  Such measurements determine the top squark's charge, mixing angle and coupling to the electroweak symmetry breaking sector.  Determining these couplings can provide strong evidence for the supersymmetric solution to the hierarchy problem.  Our analysis shows that the LHC has great prospects for measuring the photon and Higgs couplings, but will require a very high luminosity to measure the $Z$ coupling.
\end{abstract}

\maketitle

The LHC era is well underway and the next particle physics discovery could be imminent.
Determining the fundamental significance of a new particle requires measurements of its properties, some easier than others.
For example, its mass is straightforward to measure by kinematics, but determining its spin is challenging.
The interactions of the new particle, specifically what
it couples to and how strongly, are also quite difficult to measure for the following reasons.  The production of a new particle
is typically dominated by one type of coupling (e.g.~QCD interaction for a colored particle),
and it is extremely subtle to extract the dependence on subdominant couplings from
the production rate.
On the other hand, the decays of the particle can only tell us the ratios of 
different coupling strengths from branching ratios, with no information on their overall scale.

One can overcome this by considering the ``associated production'' of the new particle with
a known particle.  The method is extremely general:  given an LHC process producing
a new particle $X$, we consider the higher order diagram where the $X$ particle radiates a
Standard Model (SM) particle $Y$ via the coupling $\lambda\, XXY$.
This
associated $Y$ production has a rate proportional to $\lambda^2$, thereby allowing us to measure
$\lambda$.
For example, in the SM, measurements of the associated production of
$W + \gamma$ \cite{Abazov:2009hk,:2009us} and $t\bar{t} + h$ \cite{Aad:2009wy,Ball:2007zza, Plehn:2009rk} can be used to measure
the charge of the $W$ and the top quark's coupling to the Higgs, respectively.

In this letter, we consider such measurements in theories with supersymmetry (SUSY), specifically for
the {\em top squark} (or {\em stop}), i.e. the SUSY partner of the top quark, denoted as $\tilde{t}$.
The stop couplings to $\gamma$, $Z$, and $h$ are all potentially measurable.
The rate of associated $\tilde{t}\tilde{t}^*$ production with a photon
determines the charge of the top squark, verifying it is an up-type squark.
The $\tilde{t}\tilde{t}^*+Z$ rate determines the mixing angle between the two stop gauge eigenstates.
Finally, the associated production with a Higgs determines how strongly the top squark couples
to the source of electroweak symmetry breaking (EWSB).

Extracting these couplings from the observed rates requires knowledge of the stop masses as well
as a method of identifying the stops.  To demonstrate our procedure and factor out
coupling measurements from these issues, we consider a case where the stop is
stable on collider time scales.  Then, a stop hadronizes into a long-lived massive QCD bound state,
an {\em R-hadron} \cite{Farrar:1978xj}, where its mass can be measured by ionization and time-of-flight measurements
in the detector~\cite{Fairbairn:2006gg}.  As long as its lifetime is short on cosmological scales, the $R$-hadron is only limited by collider searches.  Moreover, 
$R$-hadrons have no irreducible SM
backgrounds and can be identified with a high efficiency.
As a further simplification we use a very minimal SUSY spectrum, namely a single stop, with all other supersymmetric particles assumed heavy enough that their production rates are small compared to the stop production. This guarantees that there are no additional sources of $Z$'s or Higgses from cascade decays.

Coupling measurements via associated production of a
new particle are general and not restricted to the top squark.  There is, however,
a deep theoretical motivation for measuring the stop couplings: it allows one to experimentally
verify that SUSY solves the hierarchy problem of the EWSB scale.
SUSY relates the stop-Higgs coupling to the top-Higgs coupling so that
the quadratically divergent contribution of the top quark to the Higgs mass-squared parameter
is precisely cancelled by that of the top squark.   Thus, the stop coupling measurements will test this cancellation that allows the EWSB scale to be naturally lower than the Planck scale.
Our method is complementary to the ``sum rule'' approach to verifying SUSY
naturalness~\cite{Blanke:2010cm} as well as the tests of other supersymmetric relations among
interactions~\cite{Allanach:2010pp, Freitas:2007fd}.
Our method can also be used to test the naturalness of Little Higgs models, in addition to other
approaches~\cite{Burdman:2002ns, Han:2003wu, Perelstein:2003wd, Han:2005ru}.

The rest of the paper is as follows: we first review the interactions of top squarks to
SM particles and choose benchmark coupling values.  We then analyze the benchmark
in detail to determine expected sensitivities of our procedure at the LHC.  Finally, we conclude.

\begin{center}
{\bf SUSY Relations and the Benchmark}
\end{center}

In this section, we discuss stop interactions with SM particles,
with the goal of determining benchmark coupling values.
In the Minimal Supersymmetric
Standard Model (MSSM), there are two gauge eigenstates for the stops,
$\tilde{t}_L$ and $\tilde{t}_R$,  SUSY partners to the SM fields, $t_L$ and $t_R$. Mass eigenstates come from diagonalizing the stop mass matrix
in the $\tilde{t}_L$-$\tilde{t}_R$ basis:%
\bea
m_{LL}^2 &=& \displaystyle{m_L^2 + m_t^2 + \left( \frac{g^2}{8} - \frac{g'^2}{24} \right) v^2 \cos2\beta}, \\
m_{LR}^2 &=& m_{RL}^2=  \displaystyle{-\left( A_t \sin\beta + y_t \mu \cos\beta \right) \frac{v}{\sqrt2}},\\
m_{RR}^2 &=& \displaystyle{m_R^2 + m_t^2 + \frac{g'^2}{6} v^2 \cos2\beta}.
\eea
The parameters, assumed to be real, are $m^2_{L,R}$, the soft SUSY breaking masses of the stops, 
the coefficient of the scalar trilinear coupling
$H_u \tilde{q}_{3L} \tilde{t}_R$, and $\mu$, the Higgsino mass parameter.
The SM parameters are the $SU(2)_L \times U(1)_Y$ couplings
$g$ and $g'$, the top quark mass $m_t$, and the EWSB vev
$v^2 = v_u^2 + v_d^2 \simeq (247 \GeV)^2$, where $v_{u,d}/\sqrt{2} \equiv \langle H^0_{u,d} \rangle$.
Finally, the SUSY top yukawa coupling $y_t$ is
fixed through $m_t = y_t v_u/\sqrt2$, and $\tan \beta \equiv v_u/v_d$.
For a review, see Ref.~\cite{Martin:1997ns}.
The mass eigenstates $\tilde{t}_{1,2}$, ordered in increasing mass,
are  $\tilde{t}_1 =   \tilde{t}_L \cos\theta_t +  \tilde{t}_R \sin\theta_t \, , \tilde{t}_2 =   \tilde{t}_L \sin\theta_t - \tilde{t}_R \cos\theta_t$.%

As mentioned earliner, we will now focus on the interactions of $\tilde{t}_1$ only, assuming that $\tilde{t}_2$
is sufficiently heavy and decoupled.   We will comment in the conclusions on the complications due to $\tilde{t}_2$.

The $\tilde{t}_1$ couplings relevant for single $\gamma/Z$ emission are:%
\bea
  \mathcal{L}_\text{gauge}
  &  =& i\!\left( \frac23 e\, A_\mu + g_{{}_{\st_1\st_1Z}} Z_\mu \right)\!
        (\partial^\mu \st_1^*) \st_1 + \mathrm{c.c.} \,, \nonumber \\
   g_{{}_{\st_1\st_1Z}} &=&
     e\biggl[ \biggl( -\frac12\cot\theta_W +\frac16 \tan\theta_W \biggr) \cos^2\theta_t
                    \\ \nonumber
                 \qquad\qquad\quad
 &&            +\frac23\tan\theta_W \sin^2\theta_t
     \biggr] \,.
\eea
Furthermore, for each of these gauge boson couplings, there is a 4-point
vertex including a gluon, which we include when we generate inclusive event samples.

The stop couplings to the Higgs are more complicated, due to the mixing of the two Higgs fields,
$H_u$ and $H_d$.  Again, for the purpose of demonstrating the principle, we 
take the lightest Higgs eigenstate $h$ to be aligned with the direction of the Higgs
vev, $(v_u, v_d)$, in the $H_u$-$H_d$ space.  This is the case in the 
decoupling limit, where the other Higgs bosons $A^0$, $H^0$, $H^\pm$ are much heavier than $h$.
The coupling of $\st_1$ to $h$ is then given by
\bea
  \mathcal{L}_\text{Higgs}
   & =& -\lambda_{{}_{\st_1\st_1h}}  h |\tilde{t}_1|^2, \nonumber \\
   \label{eqn:higgstrilinear}
   \lambda_{{}_{\st_1\st_1h}} v &=&
     2m_t^2 - \frac12 \bigl( m_{\st_2}^2 - m_{\st_1}^2 \bigr) \sin^2 2\theta_t  \\
     &+& 2v^2 \!\left[ \!\left( \frac{g^2}{8} - \frac{g'^2}{24} \right)\! \sin^2\theta_t
                    +\frac{g'^2}{6} \cos^2\theta_t
            \right]\! \cos2\beta \,.\nonumber
\eea

Here, we see that, for small mixing ($\theta_t \to 0$ or $\pi/2$) and gauge couplings
($g$, $g' \to 0$), $  \lambda_{{}_{\st_1\st_1h}} \approx 2m_t^2/v = y_t^2 v$.
This contribution comes from the quartic coupling between stops and Higgs bosons,
$y_t^2|H_u^0|^2(|t_L|^2+|t_R|^2)$, that is required by SUSY to cancel the top quark contribution to
the Higgs mass quadratic divergence.  Thus, measuring a coupling of this size would be strong evidence
that SUSY solves the naturalness problem of EWSB.
On the other hand, the mixing term in Eqn.~\ref{eqn:higgstrilinear} tends to cancel this contribution and, for large enough stop mass
differences, dominates the trilinear coupling.  In that case, testing SUSY naturalness requires
measuring both stop masses and the mixing angle in order to isolate the interesting contribution
from the Higgs-Higgs-stop-stop quartic coupling.

\begin{figure}[t]
\begin{center}
\includegraphics[width=.9\linewidth]{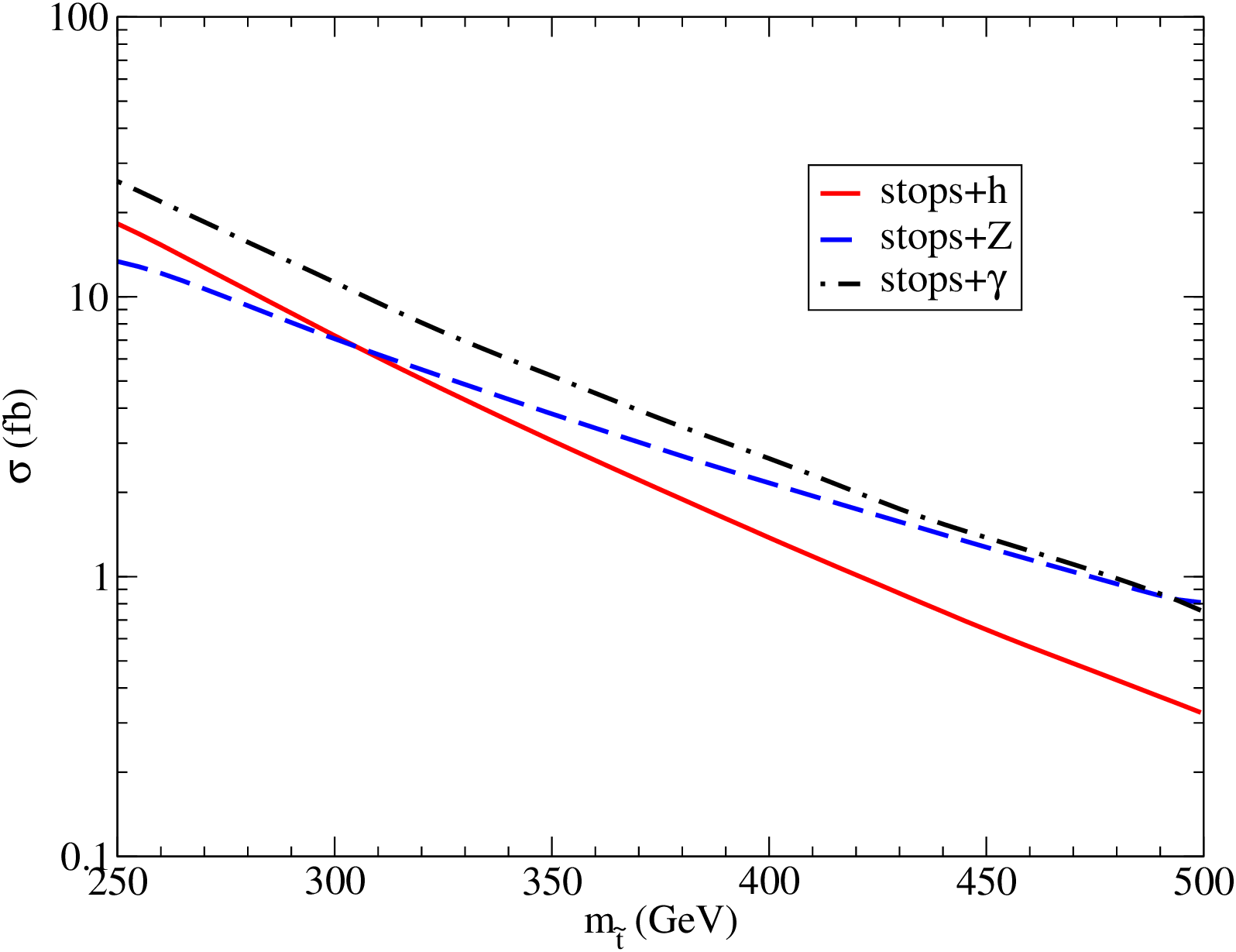}
\end{center}
\caption{The associated production cross sections as a function of stop mass with the benchmark couplings.  The contributions from initial state radiation have been subtracted.}
\label{fig:production}
\end{figure}

As a benchmark, we take the case where the lightest stop is right-handed ($\theta_t = \pi/2$).
This determines that $g_{{}_{\st_1\st_1Z}} \approx 0.11$ and
$\lambda_{{}_{\st_1\st_1h}} \approx  230\, {\rm GeV}$ for moderately large $\tan\beta$.
For other scenarios, these couplings can be larger.  For instance, the $Z$ coupling goes from $0.11$
to $-0.27$ as $\theta_t$ varies from $\pi/2$ to $0$.  The Higgs trilinear can also be much larger;
with nontrivial mixing and splitting between the stops the coupling can become an order of magnitude larger than the $2m_t^2/v$ value.  Therefore, our benchmark provides conservative estimates of
the sensitivities to the stop couplings to $Z$ and $h$. For the benchmark values, we determine the collider rates of associated production of $\gamma, Z, h$ with the $\stopR$.  See Fig.~\ref{fig:production} for the values.

\begin{center}
{\bf Collider Analysis}
\end{center}

To perform our collider analysis, we follow the recent CMS and ATLAS $R$-hadron analyses \cite{Khachatryan:2011ts, Aad:2011yf}.  Triggering is a crucial issue in an $R$-hadron search, where CMS employs muon, jet, and calorimetric missing $E_T$ (MET) triggers, while ATLAS only uses the last option.  Since $R$-hadrons are minimum-ionizing particles, they will cause an imbalance of calorimetric transverse energy. At production, $R$-hadrons can hadronize as neutral or charged and can change charge through interactions with the detector material, see \cite{Mackeprang:2009ad}.  Thus, they can appear as completely neutral with no track in the inner detector.  Even if they become charged on the way out and reach the muon system, reconstruction will likely fail without an associated track.  If instead, it hadronizes as a charged particle, reconstruction difficulties can still arise if they neutralize while traversing the detector or if they undergo charge oscillations, leading to oppositely curved tracks in the muon system. Furthermore, since $R$-hadrons are heavy, they are not ultra-relativistic and may not make their way to the outer parts of the detector within the same bunch crossing.

Due to the added subtleties of triggering on $R$-hadrons as heavy muons (which requires that the $R$-hadron retain its charge all the way through the detector), and given the high efficiency of the MET trigger used by ATLAS, we choose to implement this trigger for our analysis.  We estimate the trigger threshold for the $14~{\rm TeV}$ LHC run by doubling that used at 7 TeV \cite{Aad:2011yf}, requiring the calorimetric MET of the event (neglecting the small contribution from the $R$-hadrons) to be greater than $80~{\rm GeV}$.  Furthermore to be reconstructed, at least one of the $R$-hadrons must:
I) hadronize as a charged particle (assuming this charge is retained in the tracker), II) have a velocity in the range $0.2 < \beta\gamma < 1.5$, III) have $p_{T}>50~{\rm GeV}$ and $|\eta|<1.7$ and IV) have $\Delta R>0.5$ to the nearest jet with $E_{T}>50~{\rm GeV}$.
Such requirements are adopted from the recent ATLAS search \cite{Aad:2011yf}.
We use a subroutine supplied on the Pythia \cite{Sjostrand:2006za} webpage to determine the stop charge fractions upon hadronization and find that the probability for exactly one (two) of the $R$-hadrons to be charged to be $0.48$ ($0.33$). We assume that these probabilities are independent of the kinematics of the event. We assume that when one $R$-hadron satisfies these criteria, reducible SM backgrounds become negligible.

We generate parton level events at leading order using the usrmod functionality of MadGraph \cite{Maltoni:2002qb,Alwall:2007st} for three choices of $m_{\tilde{t}}$, at 320, 360 and 400 GeV, which are above the current limits \cite{Khachatryan:2011ts, Aad:2011yf}. For each mass point, we use the MLM matching procedure in MadGraph to generate samples of $\tilde{t}\tilde{t}^{*}+$jet(s), $\tilde{t}\tilde{t}^{*}+b\bar{b}$ (including a SM Higgs signal with $m_{h}=120~{\rm GeV}$), $\tilde{t}\tilde{t}^{*}Z$+jet and $\tilde{t}\tilde{t}^{*}\gamma$+jet. We then pass the events through Pythia for showering and hadronization and PGS to simulate detector effects. We use the default ATLAS parameter set for PGS, with a cone jet algorithm of $\Delta R=0.7$. Since stable colored particles are difficult to handle with these tools, we modify the parton level events such that the stops are treated as particles with no SM charges, and reinsert them into the event record in the offline analysis.

\begin{figure}[t]
\begin{center}
\includegraphics[width=.9\linewidth]{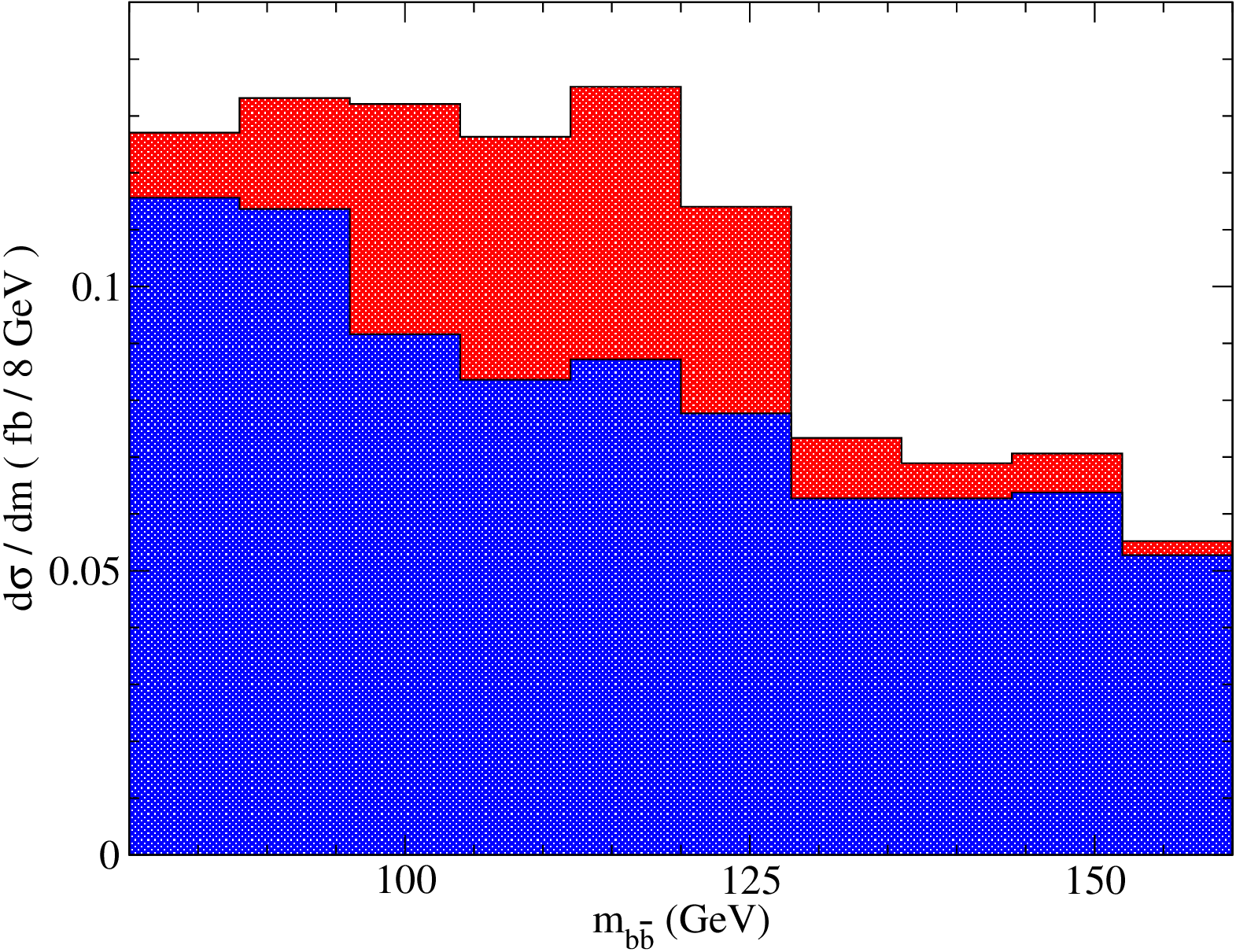}
\end{center}
\caption{The invariant mass of $b$-jet pairs in association with top squarks of $m_{\st} = 320 \GeV$.  The signal of a 120 GeV Higgs is stacked on top of the background.}
\label{fig:higgs}
\end{figure}

For the Higgs search, we use the $\tilde{t}\tilde{t}^{*}+$jet(s) sample as background and $\tilde{t}\tilde{t}^{*}+b\bar{b}$ as both signal and background since in addition to Higgs production this sample includes QCD and electroweak sources of $b$ quark production. We demand two heavy flavor tags in each event, with the following tagging procedure: for each $b$-parton in the event, we find the highest $p_T > 20 \GeV$ PGS jet, within $\Delta R=0.7$ of the parton.    We assume a flat b-tag rate of 0.6 for such jets and a 0.02 mistag rate for light jets.  

Looking at the distribution of the invariant mass of two $b$-tagged jets in the range [80,160] GeV, see for e.g.  Fig.~\ref{fig:higgs}, we chose to analyze the [96,128] GeV mass window.  For this bin, the measured coupling's $1\sigma$ error from the change in the log-likelihood is $\frac{\delta \lambda}{\lambda} = \frac{\sqrt{S+B}}{2S}.$  We require a measurement of the Higgs coupling to an accuracy of 25\%, $\frac{S}{\sqrt{S+B}}=2$, where we treat the events that have a Higgs resonance at parton level as signal. The target luminosity for this criterion can be found in Table~\ref{table:sensitivity}.

\renewcommand{\arraystretch}{1.4}
\begin{table}[t] 
\begin{center}
\begin{tabular}{|c||c|c||c||c|c|}
\hline
$m_{\st}$ & Higgs & 2$\times$Higgs & $\gamma$ & $Z$ & 2$\times Z$ \\
\hline
320 & 71 $\left(\frac{11.8}{23.1}\right)$& 9 $\left(\frac{6.0}{2.9}\right)$&6 $\left(\frac{8.3}{8.9}\right)$& 1050 $\left(\frac{14.3}{36.5}\right)$&  121$\left(\frac{6.6}{4.2}\right)$\\
\hline
360 & 135 $ \left(\frac{12.4}{26.0}\right)$& 17 $\left(\frac{6.1}{3.2}\right)$&12 $\left(\frac{9.1}{11.8}\right)$& 3600 $\left(\frac{20.1}{80.6}\right)$&  360$\left(\frac{8.0}{8.1}\right)$\\
\hline
400 & 282  $\left(\frac{13.9}{34.2}\right)$& 33 $\left(\frac{6.5}{4.0}\right)$&19 $\left(\frac{9.2}{12.0}\right)$& 4330 $\left(\frac{18.1}{63.6}\right)$&  450 $\left(\frac{7.5}{6.6}\right)$\\
\hline
\end{tabular}
\caption{Luminosity (in fb$^{-1}$), followed by $S/B$ for that luminosity in parentheses, required to measure the benchmark stop couplings  to Higgs, $\gamma$ and $Z$ with a statistical error of 25\%.  For the Higgs and $Z$, the luminosity required for a coupling twice as large is also listed.
\label{table:sensitivity}}
\end{center}

\end{table}

 In order to study the coupling of the stop to the photon, we use the $\tilde{t}\tilde{t}^{*}\gamma$+jet as signal and the $\tilde{t}\tilde{t}^{*}+$jet(s) sample as background, where we assume a flat rate of $5\times10^{-4}$ as the probability of a jet faking a photon. We also generate a $\tilde{t}\tilde{t}^{*}\gamma$+jet sample where the stops have no electric charge, i.e. the photons come from initial state radiation (ISR) only. We consider these ISR photons as part of the background. By studying another sample with the stops having twice their normal electric charge we find that the cross section of signal minus background to scale like the stop charge squared, demonstrating that photons from stops do not significantly interfere with ISR. We use the same trigger criteria as described above and accept only photons with $|\eta|<2.5$ and $p_T > 50~$GeV. We find that a second trigger (100 GeV photon + 50 GeV MET) has negligible effect on the analysis.

\begin{figure}[t]
\begin{center}
\includegraphics[width=.9\linewidth]{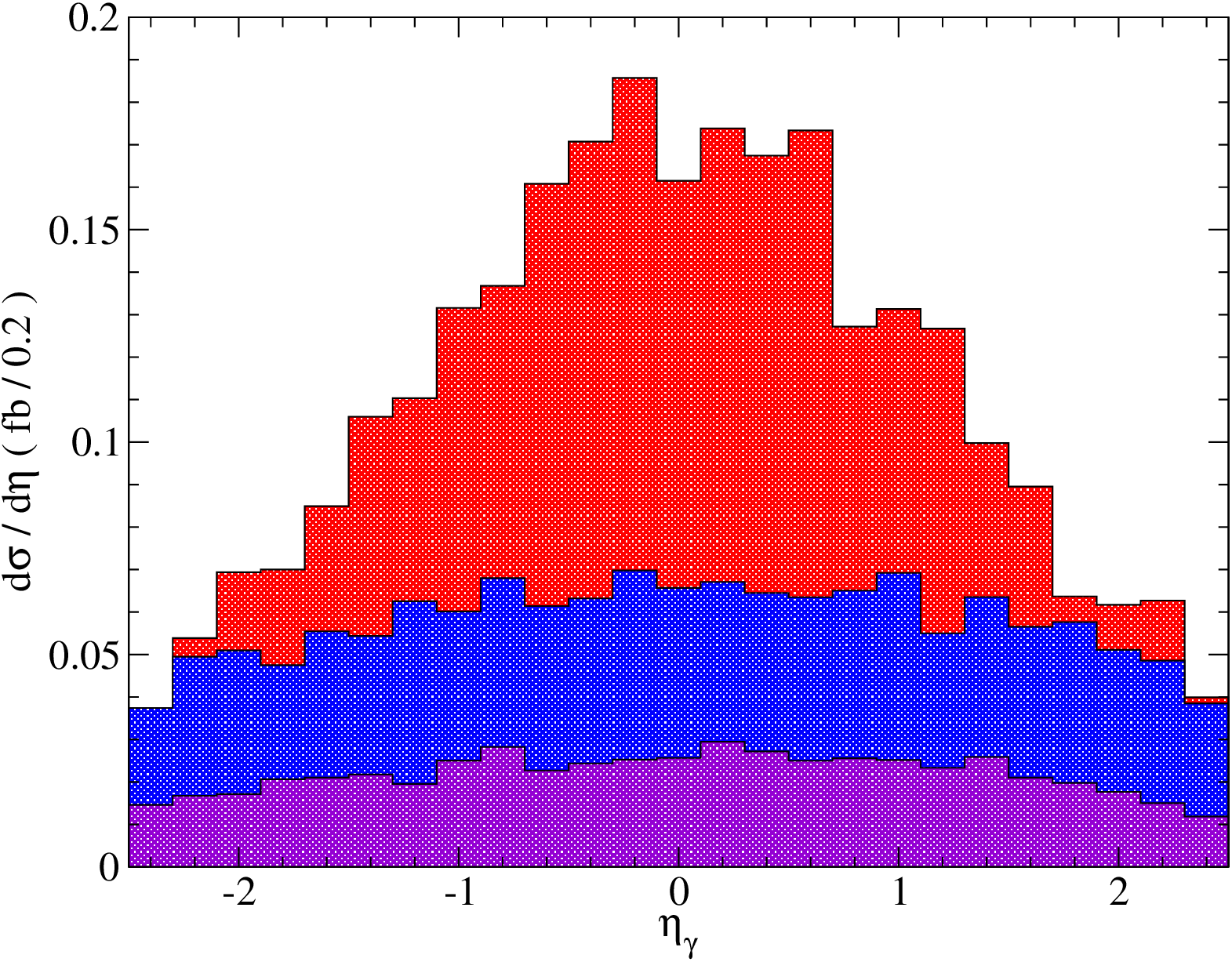}
\end{center}
\caption{The $\eta$ distribution of photons, with  $p_T > 50 \GeV$,  produced in association with top squarks of $m_{\st} = 320 \GeV$.  Stacked from top to bottom are signal, ISR photons, and the jet fake background.}
\label{fig:photon}
\end{figure}

While the $p_T$ distribution of the signal and background look very similar, the fake and ISR photons have a flat distribution in rapidity while the photons emitted from stops peak at central values and cut off at $|\eta|\sim 2$, see Fig.~\ref{fig:photon}. Therefore, in a realistic analysis, the $|\eta|>2$ could be used as a sideband, but here we take the whole range.  For the stop charge measurement, the fake and ISR photons are background, and those produced off stops are signal.  Table~\ref{table:sensitivity} lists the luminosity needed to measure the stop charge to 25\%, showing it is possible to measuring the charge for an $R$-hadron in early data.  

Finally in order to measure the $Z$ coupling to the stop, we use the $\tilde{t}\tilde{t}^{*}Z$+jet sample and consider leptonic $Z$ decays. We use a trigger based on the presence of an oppositely charged $e$ or $\mu$ pair, where both leptons have $p_{T} > 25~$GeV, $|\eta|<2.5$ and an invariant mass within $6\GeV$ of $m_Z$.   Backgrounds for this final state are assumed negligible. Unfortunately, unlike the $\gamma$ search, there is no  simple kinematic variable that distinguishes $Z$'s from ISR versus  those produced off stops. Plus, the small $Z$ branching ratio to leptons leaves little statistics  to cut on a differential distribution, forcing us to rely on Monte Carlo and do a simple counting experiment.

We treat $Z$'s from ISR as background, and as in the photon case, we verify that there is no significant interference between ISR and emission from stops. As shown in Table~\ref{table:sensitivity}, we find that an integrated luminosity of $O({\rm ab}^{-1})$ is needed to measure the $Z$ coupling with 25\% accuracy.  Prospects for a coupling twice as large, close to that of a $\tilde{t}_L$, are much more promising, but still require $\gtrsim 100 {\rm\, fb}^{-1}$.  Note that due to limited statistics, our results are more uncertain for this case than the others.

\noindent {\bf Uncertainties:} The 25\% error on these couplings is only statistical and does not take into account systematics.   The uncertainty from the  Higgs mass will be small, since it will be measured to $\sim 1\%$ in its diphoton decay \cite{Aad:2009wy,Ball:2007zza}, given $\sim 30 \invfb$.  The mass resolution of a stop $R$-hadron is expected to be $\sim12\%$ \cite{Khachatryan:2011ts} and conservatively, we assume the mass can be measured to 5\%.  Propagating this uncertainty into the signal cross section, using Fig.~\ref{fig:production}, leads to coupling uncertainties of 10-15\%.    Finally,  further uncertainties due to efficiencies and K-factors require rigorous experimental and theoretical analyses to be quantified.

\begin{center}
{\bf Concluding Remarks}
\end{center}

In this note, we have studied the LHC capabilities to measure top squark couplings to the photon, Higgs and $Z$ boson, assuming a simplified scenario of a long-lived top squark.  For conservative values of these couplings, those of a right-handed stop, we found the Higgs and photon couplings can be measured with $\lesssim 100 \invfb$, while the $Z$ coupling requires $O({\rm ab}^{-1}).$   This suggests that it is possible to identify an $R$-hadron by its charge as an up-type squark and argue that it is a top squark through its strong Higgs coupling.  Moreover, the benchmark Higgs coupling comes dominantly from the supersymmetric generalization of the top quark yukawa.  A measurement consistent with this value would be strong evidence that the top squark's couplings are supersymmetric and that supersymmetry is solving the hierarchy problem.

It is worth exploring these measurements in scenarios where the top squarks promptly decay into top quarks and a neutralino.  Since top quarks produce a bottom-jet each, there are combinatorial issues for the Higgs measurement, which could benefit from existing top squark studies \cite{Plehn:2010st, Plehn:2011tf}.  Another issue is the potential contamination from decays of the heavier top squark (or sbottom).  Such events could be distinguished by the additional cascade decay products.  In cases where these heavy particles are relevant, there are additional handles on the stop mixing angle, such as the heavy stop mass and the SUSY Higgs mass prediction.   

To conclude, this measurement technique can be applied to any new particle that has a large direct production rate at the LHC.  We hope it will become experimentally useful as the LHC era evolves.

\noindent {\bf Note added:}  During the process of this project, we learned of independent work by M.~ Luty and D.~Phalen that considers similar Higgs signals produced with $R$-hadrons \cite{Luty:2011hi}.

\noindent {\bf Acknowledgments:}  
We thank T.~Adams, J.~Chen, K.~Cranmer, Y.~Gershtein, M.~Perelstein, S.~Samolwar, and G.~Watts for useful conversations.  The work of SC is supported by DOE Grant \#DE-FG02-91ER40674 and the work of CK is supported by DOE grant \#DE-FG02-96ER40959.

\bibliographystyle{apsrev}
\bibliography{stoptrilinear}

\newpage

\end{document}